\documentclass[manuscript]{aastex61}

\newcommand\sbp{SBP}
\newcommand{\ylee}[1]{{\textcolor{black}{#1}}}
\newcommand{\yleem}[1]{{\textcolor{black}{#1}}}
\newcommand{\kms}{$\,$km$\,$s$^{-1}$}

\shortauthors{Environmental variations of SBP}

\begin{document}
\title{Difference in dwarf galaxy surface brightness profiles as a function of environment\footnote{\ylee{Based on data collected at KMTNet Telescope and SDSS}}}

\correspondingauthor{Youngdae Lee}
\email{ylee@kasi.re.kr}

\author{Youngdae Lee}
\affil{Korea Astronomy and Space Science Institute, 776, Daedeokdae-ro, Yuseong-gu, Daejeon 34055, Republic of Korea}

\author{Hong Soo Park}
\affil{Korea Astronomy and Space Science Institute, 776, Daedeokdae-ro, Yuseong-gu, Daejeon 34055, Republic of Korea}
\affil{Korea University of Science and Technology (UST), 217 Gajeong-ro, Yuseong-gu, Daejeon 34113, Republic of Korea}

\author{Sang Chul Kim}
\affil{Korea Astronomy and Space Science Institute, 776, Daedeokdae-ro, Yuseong-gu, Daejeon 34055, Republic of Korea}
\affil{Korea University of Science and Technology (UST), 217 Gajeong-ro, Yuseong-gu, Daejeon 34113, Republic of Korea}

\author{Dae-Sik Moon}
\affil{Department of Astronomy and Astrophysics, University of Toronto, St. George, Toronto, ON M5S 3H4, Canada}

\author{Jae-Joon Lee}
\affil{Korea Astronomy and Space Science Institute, 776, Daedeokdae-ro, Yuseong-gu, Daejeon 34055, Republic of Korea}

\author{Dong-Jin Kim}
\affil{Korea Astronomy and Space Science Institute, 776, Daedeokdae-ro, Yuseong-gu, Daejeon 34055, Republic of Korea}

\author{Sang-Mok Cha}
\affil{Korea Astronomy and Space Science Institute, 776, Daedeokdae-ro, Yuseong-gu, Daejeon 34055, Republic of Korea}
\affil{School of Space Research, Kyung Hee University, Yongin 17104, Korea}

\begin{abstract}

We investigate surface brightness profiles ({\sbp}s) of dwarf galaxies in \ylee{field, group, and cluster environments}. With deep $BVI$ images from the Korea Microlensing Telescope Network Supernova Program, {\sbp}s of 38 dwarfs in the NGC 2784 group are fitted by a single exponential or double exponential model. We find that 53\% of the dwarfs are fitted with single exponential profiles (``Type I"), while 47\% of the dwarfs show double exponential profiles. 37\% of all dwarfs have smaller sizes of the outer part than inner part (``Type II"), while 10\% have a larger sized outer than inner part (``Type III"). We compare these results with those in the field and in the Virgo cluster, \ylee{where {\sbp} types of 102 field dwarfs are compiled from a previous study and {\sbp} types of 375 cluster dwarfs are measured using SDSS $r$-band images. As a result, the distributions of {\sbp} types are different in three environments. Common {\sbp} types of the field, the NGC 2784 group, and the Virgo cluster are Type II, Type I and II, and Type I and III profiles, respectively. After comparing sizes of dwarfs in different environments, we suggest that since sizes of some dwarfs are changed due to the environmental effects, {\sbp} types are able to be transformed and the distributions of {\sbp} types in three environments are different. We discuss possible environmental mechanisms on the transformation of {\sbp} types.}

\end{abstract}


\section{Introduction} \label{sec:intro}
Surface brightness profiles ({\sbp}s) of dwarf galaxies are an important tool in understanding the evolution of dwarfs \citep{Kor85,Bos08a,Kor12,Mey14,You14}. With the assumption that dwarfs have a single component {\sbp}, such as King, exponential, and S\'ersic profiles, it is shown that dwarfs have size-magnitude relations where the brighter absolute magnitudes correspond with larger size \citep{Kor85,Bos08a,You14}.

In recent studies of the {\sbp}s of dwarfs, many dwarfs are shown to have two-component {\sbp}s \citep{Agu05,McD11,Her13,Jan12,Jan14,Jan16,Mey14}. The {\sbp}s are described by S\'ersic profile for the inner part and exponential one for the outer part \citep{Agu05,Jan12,Jan14,Jan16}. For the inner part, since the S\'ersic index ($n$) is approximately 1 \citep[$n<1.2$;][]{Jan14}, the inner profiles could also be described by the exponential model. Therefore, the two-component {\sbp}s of dwarfs could be fitted by double exponential {\sbp}s for both the inner and outer parts. Noteworthily, \citet{Her13} used the double exponential model to explain the {\sbp}s of dwarfs and they classified three {\sbp} types. \ylee{Type I is for classical dwarfs with single exponential profiles \citep{Fre70}. Type II \citep{Fre70} and Type III \citep{Erw05} are for double exponentials with smaller and larger outer sizes than inner sizes, respectively.}

Sizes of dwarfs measured from two-component {\sbp}s show size-magnitude relations \citep{Her13,Jan16} that vary with environments \citep{Jan16,You14}. According to \citet{Jan16}, the sizes of cluster early-type dwarfs are smaller than those of field late-type dwarfs at a given magnitude. Based on the fact that cluster early-type dwarfs are transformed from field late-type dwarfs, \citet{Jan16} claimed that cluster environmental effects (e.g., ram pressure stripping and harassment) could reduce sizes of dwarfs infalling into the cluster. In terms of {\sbp}s depending on environments, groups of galaxies are an important piece together with the field and the clusters. Groups of galaxies locate themselves between the field and clusters according to influences of the diverse environments. The galaxies in groups might have unique properties compared to those in the field and in clusters.

In this paper, we investigate the {\sbp}s of dwarfs in the NGC 2784 group observed using the Korea Microlensing Telescope Network \citep[KMTNet;][]{Kim16} Supernova Program \citep[KSP;][]{Moo16,Par17} and compare their properties with those of field and cluster galaxies. \ylee{The NGC 2784 group is defined in several studies \citep{Tul87,Fou92,Gar93,Mak11}. \citet{Tul87} measured $B$-band total luminosity ($1.905\times10^{10}\,$L$_{\odot}$), barycentric group velocity (519 \kms{}), velocity dispersion (68 \kms{}), virial radius (0.31 $h_{75}^{-1}\,$Mpc), and virial mass ($3.048\times10^{12}\,$M$_{\odot}$) for the group. \yleem{The distance (9.8 Mpc; $m-M=29.96$ mag) of the NGC 2784 galaxy is measured by surface brightness fluctuation method \citep{Ton01}, which is adopted as a distance of the NGC 2784 group in this study.} Faint X-ray source (2RXS J091217.6-241015) is detected near the center of the group \citep{Bol16}. However, it is not known yet whether the source is from intragroup medium of the group or hot galactic halo of NGC 2784. The NGC 2784 group has some bright galaxies. The three brightest members are NGC 2784 ($M_V = -20.85$ mag), N2835 ($M_V = -20.44$ mag), and ESO 565-1 (DDO062; $M_V = -17.83$ mag). The number of members are 2 to 5 galaxies depending on group defining method and sample sizes \citep{Tul87,Fou92,Gar93,Mak11}. Recently, total number of member galaxies in the NGC 2784 group is 41 galaxies including newly found 31 dwarf galaxies ($M_V < -9$ mag) from \citet{Par17}. This group shows mass segregation where massive galaxies are concentrated on the group center than less massive galaxies. Furthermore, the total $K$-band luminosity (log$L_K = 10.91$ $L_{\odot}$) and the faint-end slope ($\alpha = -1.33$) of the luminosity function of the NGC 2784 group are similar to those of the M81, M83 and M106 groups \citep{Par17}.}

In section~\ref{sec:data}, we describe how to measure the {\sbp}s of dwarfs and classify them using double exponential analysis. In section~\ref{sec:result}, we introduce the fractions and \ylee{size distributions} of group dwarfs with three {\sbp} types. The results for the NGC 2784 group are compared with those in the field and those in clusters. In section~\ref{sec:summary}, we discuss the {\sbp} \ylee{types} of dwarfs depending on environments.

\section{Data and Analysis} \label{sec:data}
To investigate reliable {\sbp}s of dwarfs in a group, deep images are required. Recently, in the nearby NGC 2784 group, 31 new dwarfs were discovered in $BVI$ deep images created with total exposure time exceeding 20000 seconds from KSP. The deep images enable us to measure reliable {\sbp}s down to $\mu_V \sim 28.5$ mag arcsec$^{-2}$ at the  3-sigma confidence level. Details of observations and data reduction are shown in \citet{Par17}. 

A total of 38 {\sbp}s of dwarfs, including seven previously known dwarfs in the NGC 2784 group, are measured using the IRAF/$ELLIPSE$ task. We only use $V$-band images for our analyses because the photometric calibration is more stable than in the other bands ($B$ and $I$). After cropping the images around the target galaxies, we mask foreground stars and bleedings due to nearby saturated stars. In order to trace the underlying {\sbp} to the outer region of a galaxy and reducing the substructures (e.g., bar or lens) in a more efficient way, we fix the ellipse center, the ellipticity, and the position angle \citep{Erw08}. The initial parameters are derived at the effective radius \ylee{($r_e$)} of each target when $ELLIPSE$ was executed with the parameters being allowed to vary. \ylee{The effective radius is measured from one component S\'ersic fit with {\sbp} data points between seeing radius and a radius of 3-sigma above the local background level. The local background levels and their sky noises are, respectively, {\sbp} values and their error at a certain {\sbp} data bin with minimum flux gradient of the smoothed growth curve.} 

For reliable classification of {\sbp}s, we only use {\sbp} data points above 3-sigma from the background level and within \ylee{3$r_e$} from the galaxy center. We then fit the {\sbp} using the exponential function of   
\begin{equation}
\mu(R) = \mu_0 + \frac{2.5}{ln(10)}\left(\frac{R}{h}\right),
\end{equation}
where $\mu_0$ and $h$ are the central surface brightness and the scale length, respectively. When the $\chi^2_{\nu}$ (reduced chi-square) is less than 2.5, profile types are assigned as a single exponential profile (Type I). When it  is larger than 2.5, profile types are assigned as a two-component profile. In establishing the criterion of $\chi^2_{\nu}=2.5$, we classify the {\sbp} visually and compare the visually classified {\sbp} types with the $\chi^2_{\nu}$-based {\sbp} types. As a result, the value of 2.5 reconciles two classifications better than any other $\chi^2_{\nu}$. 

\begin{figure*} 
\plotone{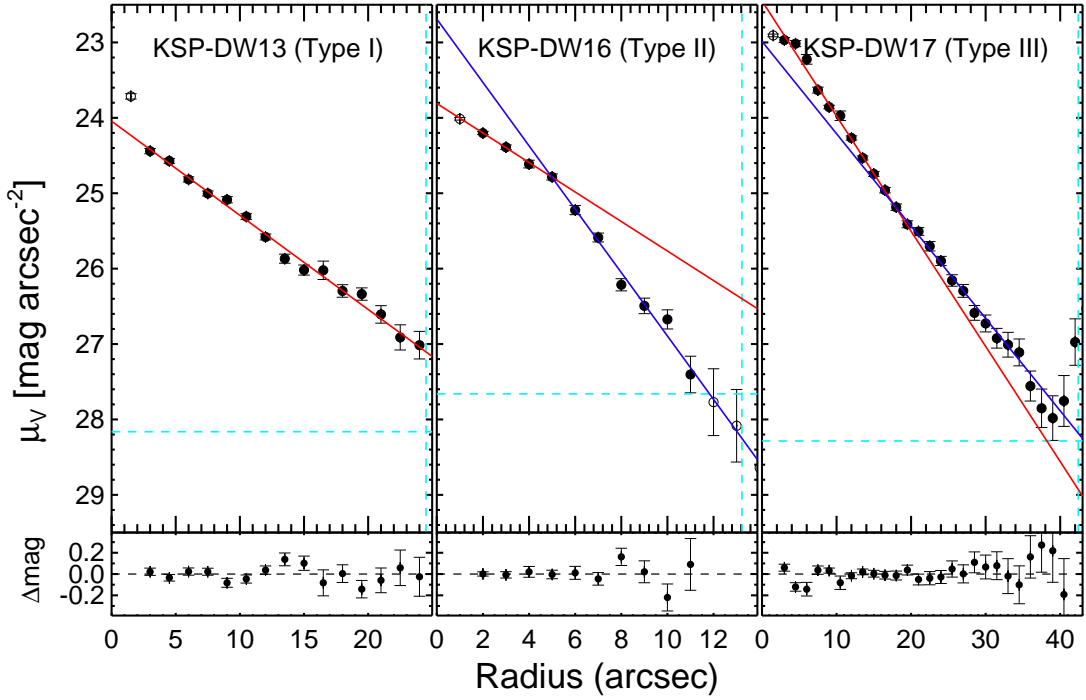}
\caption{Examples for three {\sbp} types. (Left) Type I (single exponential). (Middle) Type II (outer down-bending exponential). (Right) Type III (outer up-bending exponential). The circles are {\sbp}s from $ELLPISE$ and filled circles are points used for the double exponential fit. The red solid lines are single exponential or inner exponential models, and the blue solid lines are outer exponential models. The cyan horizontal and vertical dashed lines are 3-sigma above the background level and 3$r_{e}$ from the galaxy center, respectively. The bottom panels present the differences between the observed {\sbp}s and the best double exponential fits. \label{fig1:extype}} 
\end{figure*}

Since two-component profiles of dwarfs have been explained by double exponentials \citep{Her13,Jan14}, the profiles are divided into outer down-bending (Type II) and up-bending (Type III) profiles using a double exponential model. We follow the fitting procedure of \citet{Her13} in our double exponential fitting as follows. \ylee{The SBP is iteratively divided into every possible pair of subregions with at least two SBP data points moving from seeing radius to 3$r_e$. Then, the pair of subregions is fitted by two separate single exponentials. As the best double exponential model, we adopt a model which minimizes the sum of $\chi^2_{\nu}$s for the pair of subregions.} We consider that {\sbp}s are Type II when profiles show a smaller outer scale length ($h_o$) than the inner scale length ($h_i$) and Type III when $h_o$ is larger than $h_i$. Figure~\ref{fig1:extype} shows examples of the three types of {\sbp}s. 

In addition to the classification of {\sbp}s, we carry out morphological classification of the dwarfs via visual inspections of $BVI$ images. We divide their morphologies into early- and late-type dwarfs. Main features of early-type dwarfs are elliptical isophotes and smooth {\sbp}s, alongside well-defined positions of the galaxy centers. In contrast, late-type dwarfs show the following features : (1) irregular morphologies, (2) many star forming regions, and (3) large nuclear offsets from the galaxy center \citep{San84,Bin85,Kim14}. As a result, we have 21 early-type and 16 late-type dwarfs in our data. \ylee{The morphology of one galaxy (KSP-DW6) cannot be determined due to many bleedings from nearby bright stars.} It is well known that early- and late-type dwarfs typically have red and blue colors, respectively \citep{Bos97,Jan00,Gav10,Fer12,Pak14}. Our morphological classification is in accordance with the expectation of this morphology-color relation. Mean $B-V$ colors of our early- and late-type dwarfs are 0.73 and 0.57 mag, respectively. The properties of dwarfs in the NGC 2784 group are shown in Table~\ref{tab:dwarf}.

\input{table1.tab}

\section{Results} \label{sec:result}

\subsection{Dwarf galaxies in the NGC 2784 group} \label{ssec:frequency}
Table~\ref{tab:fraction} contains the fractions and their \ylee{68.3\% confidence intervals \citep{Wil27}} for different {\sbp} types of dwarfs in the NGC 2784 group. For all morphologies, \ylee{Type I (52.6$_{-8.1}^{+7.9}$\%) and Type II (36.8$_{-7.4}^{+8.1}$\%) are more common than} Type III (10.5$_{-4.0}^{+6.0}$\%). According to \citet{Par17}, six out of 38 dwarfs may not be associated with the NGC 2784 group considering that they are located further ($>$2.7 degrees) from the group center. When we select the 32 dwarfs within 2.7 degrees from the group center, fractions of SBP types do not change significantly, i.e., 50.0$_{-8.7}^{+8.7}$\%, 40.6$_{-8.3}^{+8.8}$\%, and 9.4$_{-4.0}^{+6.5}$\% for Type I, II, and III profiles, respectively. Thus, the uncertainty of membership does not appear to seriously affect our results. Type I (38.1$_{-9.8}^{+10.9}$\%) and Type II (42.9$_{-10.2}^{+10.9}$\%) profiles seem to be equally common in the case of early-type dwarfs and a majority (68.8$_{-12.4}^{+10.2}$\%) of late-type dwarfs seem to exhibit Type I profiles. \ylee{However, the distribution of {\sbp} types between early- and late-type dwarfs are not significantly different. For this case, the $3\times2$ Fisher's exact test\footnote{We use the $R$ statistical package (\url{http://www.r-project.org/})} provides $P_F=0.107$}.

\begin{table} 
\renewcommand{\thetable}{\arabic{table}}
\centering
\caption{Fractions of {\sbp} types} \label{tab:fraction}
\begin{tabular}{c c c c}
\tablewidth{0pt}
\hline
\hline
\multicolumn{4}{c}{NGC 2784 group} \\
\hline
Morphology & Type I & Type II & Type III \\
\hline
\decimals
All (38)  &  20 (52.6$_{-8.1}^{+7.9}$\%)    & 14 (36.8$_{-7.4}^{+8.1}$\%) & 4 (10.5$_{-4.0}^{+6.0}$\%)     \\
Early-type (21) &  8 (38.1$_{-9.8}^{+10.9}$\%)   &  9 (42.9$_{-10.2}^{+10.9}$\%)  & 4 (19.0$_{-7.1}^{+9.9}$\%)     \\
Late-type (16) &  11 (68.8$_{-12.4}^{+10.2}$\%)   &   5 (31.3$_{-10.2}^{+12.4}$\%)   & 0 (0.0$_{-0.0}^{+5.9}$\%)  \\
Unknown (1) &  1 (100.0$_{-50.0}^{+0.0}$\%)   &   0 (0$_{-0.0}^{+50.0}$\%) & 0 (0$_{-0.0}^{+50.0}$\%) \\
\hline

\multicolumn{4}{c}{} \\
\hline
\hline
\multicolumn{4}{c}{Field$^{a}$} \\
\hline
Late-type (102) & 12 (11.8$_{-2.8}^{+3.6}$\%) &  70$^b$ (68.6$_{-4.8}^{+4.4}$\%) &  20 (19.6$_{-3.6}^{+4.2}$\%) \\
\hline

\multicolumn{4}{c}{} \\
\hline
\hline
\multicolumn{4}{c}{Virgo cluster} \\
\hline
All (375) & 193 (51.5$_{-2.6}^{+2.6}$\%) & 47 (12.5$_{-1.6}^{+1.8}$\%) & 135 (36.0$_{-2.4}^{+2.5}$\%) \\
Early-type (300) & 157 (52.3$_{-2.9}^{+2.9}$\%) & 23 (7.7$_{-1.4}^{+1.7}$\%) & 120 (40.0$_{-2.8}^{+2.9}$\%) \\
Late-type (75) & 36 (48.0$_{-5.7}^{+5.8}$\%) & 24 (32.0$_{-5.1}^{+5.6}$\%) & 15 (20.0$_{-4.2}^{+5.0}$\%) \\
\hline

\multicolumn{4}{c}{\ylee{NOTE.-- $^a$See Section~\ref{ssec:typespec} for field galaxy samples.}} \\
\multicolumn{4}{c}{\ylee{$^b$Field Type II includes pure Type II (58 galaxies) and Type FI (12 galaxies).}} \\
\end{tabular}
\end{table}

\begin{figure*} 
\plotone{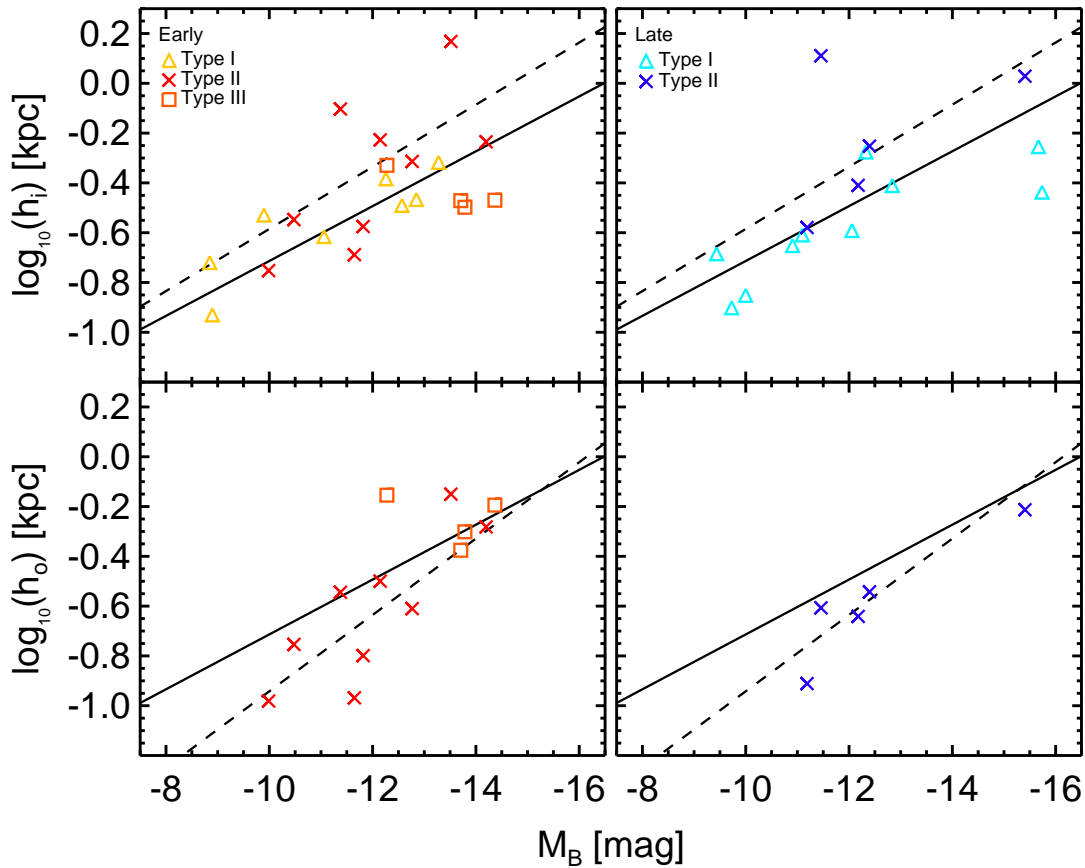}
\caption{Sizes of galaxies in the NGC 2784 group as a function of $M_B$. The top and bottom panels are for the inner and outer components of galaxies, respectively. The left panels are for early-type dwarfs and the right panels are for late-type dwarfs. {\sbp} types are shown in the legend of the top panels. The solid and dashed lines are a linear least-square fit for Type I dwarfs with $M_B>-14$ mag and for Type II dwarfs, respectively. \label{fig2:sizemag}}
\end{figure*}

In Figure~\ref{fig2:sizemag}, we investigate size-magnitude relations in which the size represents scale length. For this analysis, $B$-band magnitudes \ylee{and their foreground Galactic extinction ($A_B=0.761$)} are adopted from \citet{Par17}. Since they provide $B$-band magnitudes for only 36 dwarfs, two dwarfs (KSP-DW6 and KSP-DW23) are not used in this analysis. To characterize the size-magnitude relations, we use the linear least-square fit to dwarfs of $M_B>-14$ mag. In Figure~\ref{fig2:sizemag}, the solid and dashed lines are the measured relations for Type I and Type II dwarfs, respectively. \ylee{These size-magnitude relations are followings :}
\begin{equation}
log_{10}(h_i) = -0.1102\times M_B -1.8152 \qquad\textrm{for size of Type I} 
\end{equation}
\begin{equation}
log_{10}(h_i) = -0.1248\times M_B -1.8326 \qquad\textrm{for inner size of Type II}
\end{equation}
\begin{equation}
log_{10}(h_o) = -0.1534\times M_B -2.4756 \qquad\textrm{for outer size of Type II}
\end{equation}

We compare distributions of sizes between Type I and Type II profiles, but Type III profiles are not analyzed due to their small populations. \ylee{Since these relations show increasing sizes with increasing luminosity, we actually compare sizes normalized by the size-magnitude relation of Type I (Equation 2). As a result, the distribution of inner sizes\footnote{In this study, sizes of Type I profiles are indicated as inner or outer sizes because they are the same.} of Type I profiles is similar to that of Type II profiles, in which the one dimensional two sample Kolmogorov-Smirnov (K-S) test suggests that the sizes of Type I and Type II profiles are drawn from the same samples with a probability ($P_{KS}$) of 0.055. The distribution of outer sizes of Type I profiles is also similar to that of Type II profiles ($P_{KS}=0.093$). In addition, the distributions of sizes for early- (left panels) and late-type (right panels) dwarfs are compared. Sizes between early- and late-type Type I dwarfs show no distinct distributions ($P_{KS}=0.148$). Inner sizes (upper panels) of Type II profiles between them show $P_{KS}=0.779$ and for outer sizes (bottom panels), $P_{KS}$ is 0.425.}

\begin{figure*} 
\plotone{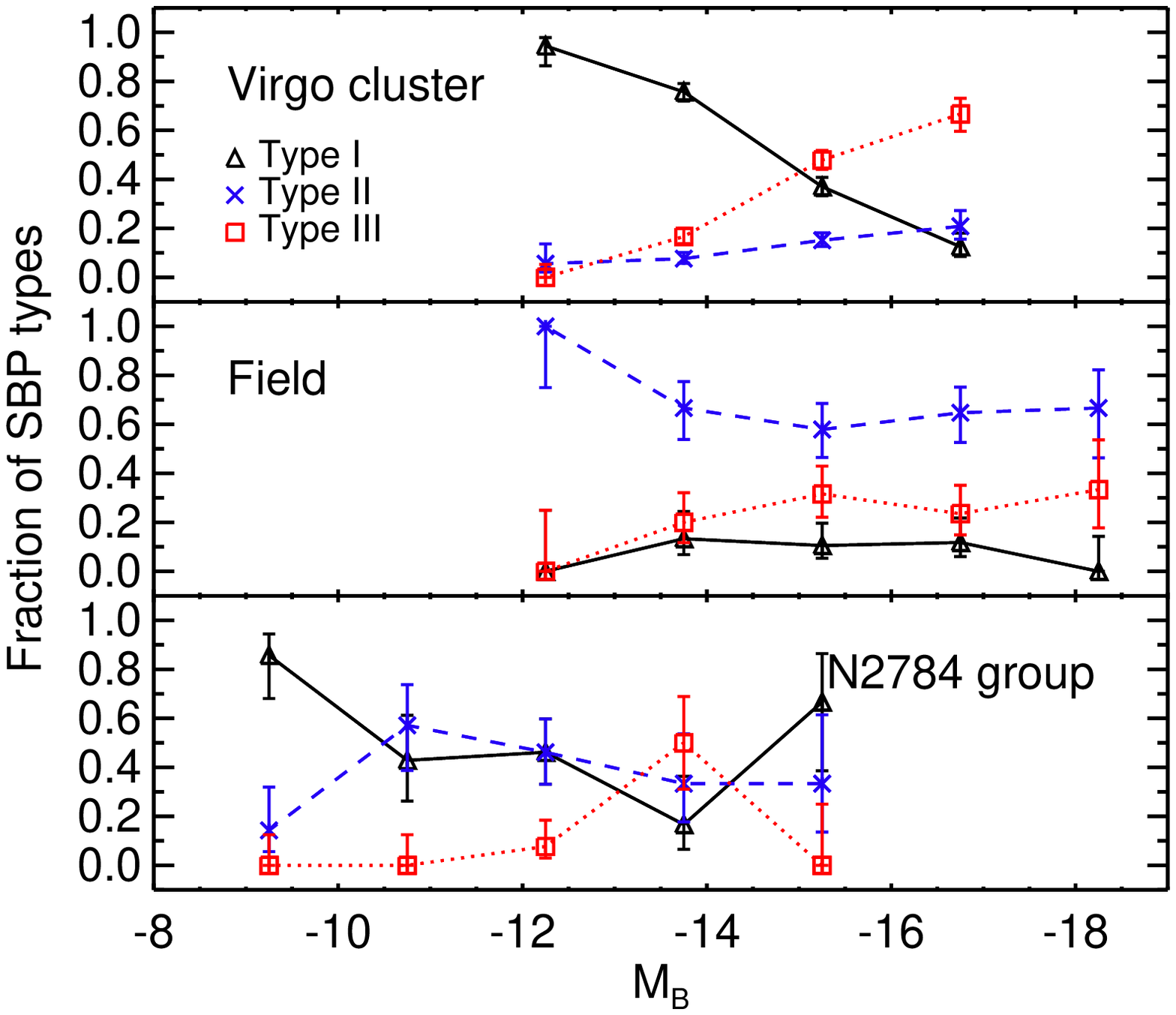}
\caption{{\sbp} fractions as a function of $M_B$. The Virgo cluster, field, and NGC 2784 group samples are presented in top, middle, and bottom panels, respectively. The legend is shown at the left corner in the top panel. \label{fig3:magdepen}}
\end{figure*}

\subsection{Environmental effects on the {\sbp} type fraction and \ylee{size distribution}} \label{ssec:typespec}

\subsubsection{\ylee{Field and cluster dwarfs}}
To analyze environmental effects on the {\sbp} types in dwarfs, we compare the fractions of {\sbp} types that we found in the NGC 2784 group in this work with those in the field and the Virgo cluster. Data for field dwarfs are adopted from \citet{Her13}, who used the same photometric band and {\sbp} type classification method as ours. \ylee{\citet{Her13} claimed that their 141 irregular galaxies coming from \citet{Hun04,Hun06} are relatively isolated in terms of tidal indices (see their section 4.5). However, these galaxies could be located on the outskirt of groups with various sizes. To verify whether they are group or field galaxies, we take advantage of a group catalog created by \citet{Mak11} which includes 395 groups identified among 10914 galaxies with $|b| > 15$ degree and $V_{LG} < 3500$\kms{}. Out of Herrmann's 141 galaxies, 130 galaxies are included in the sample area of \citet{Mak11}. When spatial and velocity separations between the Herrmann's galaxies and the Makarov's group centers are less than $R_{max}$ and $V_{max}$, Herrmann's galaxies are assumed as group galaxies. $R_{max}$ and $V_{max}$ are maximum spatial and velocity separations of Makarov's group members from the group center. The other galaxies in Herrmann's catalog that are not considered as group galaxies are defined as field galaxies. As a result, 28 and 102 Herrmann's galaxies are defined as group and field galaxies, respectively. About 78\% (102/130) of Herrmann's galaxies are field galaxies. All the field dwarfs are late-type galaxies. We directly make use of the {\sbp} type classifications in $V$-band, sizes, and absolute $B$-band magnitudes ($M_B$s) by \citet{Her13}.} 

\ylee{For cluster dwarf samples, we used 680 dwarfs (dE, dS0, Im, Sm, and BCD) with $-17$ mag $< M_B < -12$ mag and $v_h < 3500$\kms{} in the Virgo Cluster Catalog \citep{Bin85}. Morphological classifications are adopted from \citet{Bin85} and absolute $B$-band magnitudes are acquired from the HyperLeda catalog \citep{Pat03}, assuming $(m-M)=31.09$ mag \citep{Mei07}. Heliocentric radial velocities are from \citet{Kim14}. Among the selected 680 dwarf samples, we remove 305 dwarf samples which are possibly contaminated by halos of nearby bright galaxies or stars in SDSS $r$-band images. Finally, 375 dwarf samples are secured to analyze {\sbp}s, which are composed of 300 early-type (dE, dS0) and 75 late-type (Im, Sm, BCD) dwarf galaxies. From SDSS $r$-band images for the 375 dwarf galaxies, we measured {\sbp}s and classified the {\sbp} types via the same manner in Section~\ref{sec:data}.} 

\ylee{We compared our {\sbp} types with those derived from \citet{Roe12}. Out of 10 common targets, 7 dwarfs (VCC0620, VCC1410, VCC1654, VCC1684, VCC1890, VCC1897, VCC1952) have the same {\sbp} types, while 3 dwarfs (VCC0510, VCC1021, VCC2012) show different {\sbp} classifications. Although \citet{Roe12} does not provide the SBPs of galaxies, VCC0510 and VCC1021 are shown in their figure 3 and 4, respectively. VCC0510 and VCC1021 have Type III and Type I profiles in our classifications, respectively, while they has Type I (VCC0510) and Type II (VCC1021) profiles in \citet{Roe12}'s classifications. When we inspect their figure 3 and 4, the two dwarfs seem to have breaks at around $r_e$. \yleem{Considering the classification via visual inspections, our $\chi_{\nu}^{2}$ based classification of VCC0510 seems to be more plausible than \citet{Roe12}'s classification, while \citet{Roe12}'s classification of VCC1021 seems to be more reliable than ours.} Although some SBP types are mismatched, \citet{Roe12}'s and our classifications agree quite well. The properties of dwarf galaxies in the Virgo cluster are provided in Table~\ref{tab:virgo}.}

\input{table3.tab}

\subsubsection{\ylee{Environmental effects on the {\sbp} type fraction}}

\ylee{Table~\ref{tab:fraction} shows the fractions of {\sbp} types in the field and in the Virgo cluster as well as those in the NGC 2784 group. The distributions of {\sbp} types in different environments are \yleem{quite} different for all mophological types. The $3\times2$ Fisher's exact test provides $P_F=5.23\times10^{-6}$ for the group and field galaxies. In the case of the comparison between the group and cluster samples and between the field and the cluster samples, $P_F$s are $6.71\times10^{-5}$ and $2.2\times10^{-16}$, respectively. Interestingly, we find that common {\sbp} types in each environment are also different. Type II (68.6$_{-4.8}^{+4.4}$\%), Type I (52.6$_{-8.1}^{+7.9}$\%) and II (36.8$_{-7.4}^{+8.1}$\%), and Type I (51.5$_{-2.6}^{+2.6}$\%) and III (36.0$_{-2.4}^{+2.5}$\%) profiles  are common in the field, the NGC 2784 group, and  in the Virgo cluster, respectively. In the case of early-type dwarfs, the {\sbp} type distribution of group galaxies are \yleem{quite} different from that of cluster early-type dwarfs ($P_F=8.55\times10^{-5}$). In contrast, the late-type dwarfs in the Virgo cluster are the similar distribution to that of group late-type dwarfs ($P_F=0.095$). Field late-type galaxies have a different {\sbp} type distribution to that of group late-type galaxies ($P_F=5.60\times10^{-6}$) and to that of cluster late-type galaxies ($P_F=8.96\times10^{-8}$).}

\ylee{For different mophological types, early- and late-type dwarfs in the NGC 2784 group show similar {\sbp} type distributions (see Section~\ref{ssec:frequency}), while in the Virgo cluster, early- and late-type dwarfs have different {\sbp} type distributions ($P_F=2.31\times10^{-7}$).}

Since magnitude ranges of dwarf samples in each environment are different, these differences may be the source of our findings. To check dependences on magnitudes, we investigate the fractions of {\sbp} types as a function of magnitude in Figure~\ref{fig3:magdepen}. \ylee{In the top panel, we found that the {\sbp} type fractions of dwarfs in the Virgo cluster vary depending on $M_B$. The fractions of Type I and Type III significantly increase with decreasing luminosities and with increasing luminosities, respectively. However, the fractions of {\sbp} types in the field and the NGC 2784 group show no strong dependency on magnitude in the middle and bottom panels. These results imply that although the fractions of {\sbp} types in the NGC 2784 group and field could be compared with no concern of magnitude ranges, comparing with the fractions of cluster dwarfs must consider magnitude ranges of samples.}

\ylee{Therefore, we compare the {\sbp} type fractions of dwarfs with $-16$ mag $< M_B < -12$ mag in different environments. Samples with $-16$ mag $< M_B <$ $-12$ mag include 20 galaxies for the NGC 2784 group, 36 galaxies for the field and 327 dwarfs for the Virgo cluster. The fractions of these samples are shown in Table~\ref{tab:fraction2}. The results of comparisons for the distributions in different environments are the same as those with the all samples (see Table~\ref{tab:Pfs}). This implies that although we used the samples with different magnitude ranges, the results are not seriously altered by differences in magnitude ranges.}

\subsubsection{\ylee{Environmental effects on the size distribution}}

\begin{figure} 
\plotone{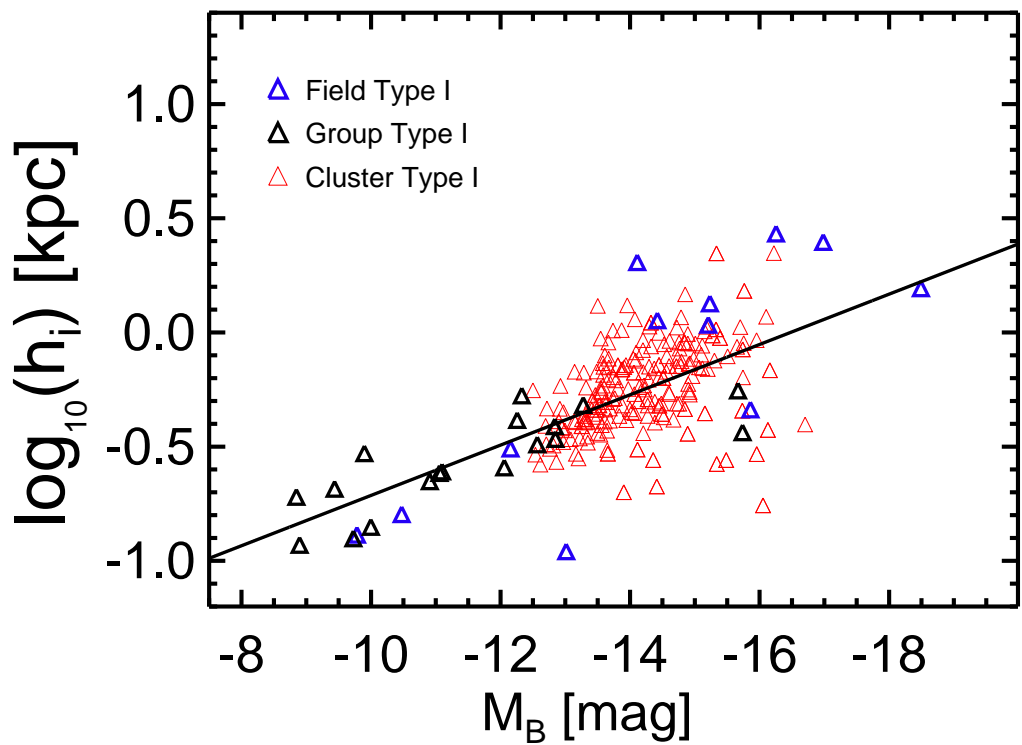}
\caption{Size distributions of Type I dwarfs as a function of environment. The dwarf galaxies in the field, group, and cluster are shown as blue, black, and red symbols, respectively. The solid line is the same solid line in Figure~\ref{fig2:sizemag}, which is measured from sizes of Type I dwarfs in the NGC 2784 group. \label{fig4:sizemagcomp_t1}}
\end{figure}

\begin{figure} 
\plotone{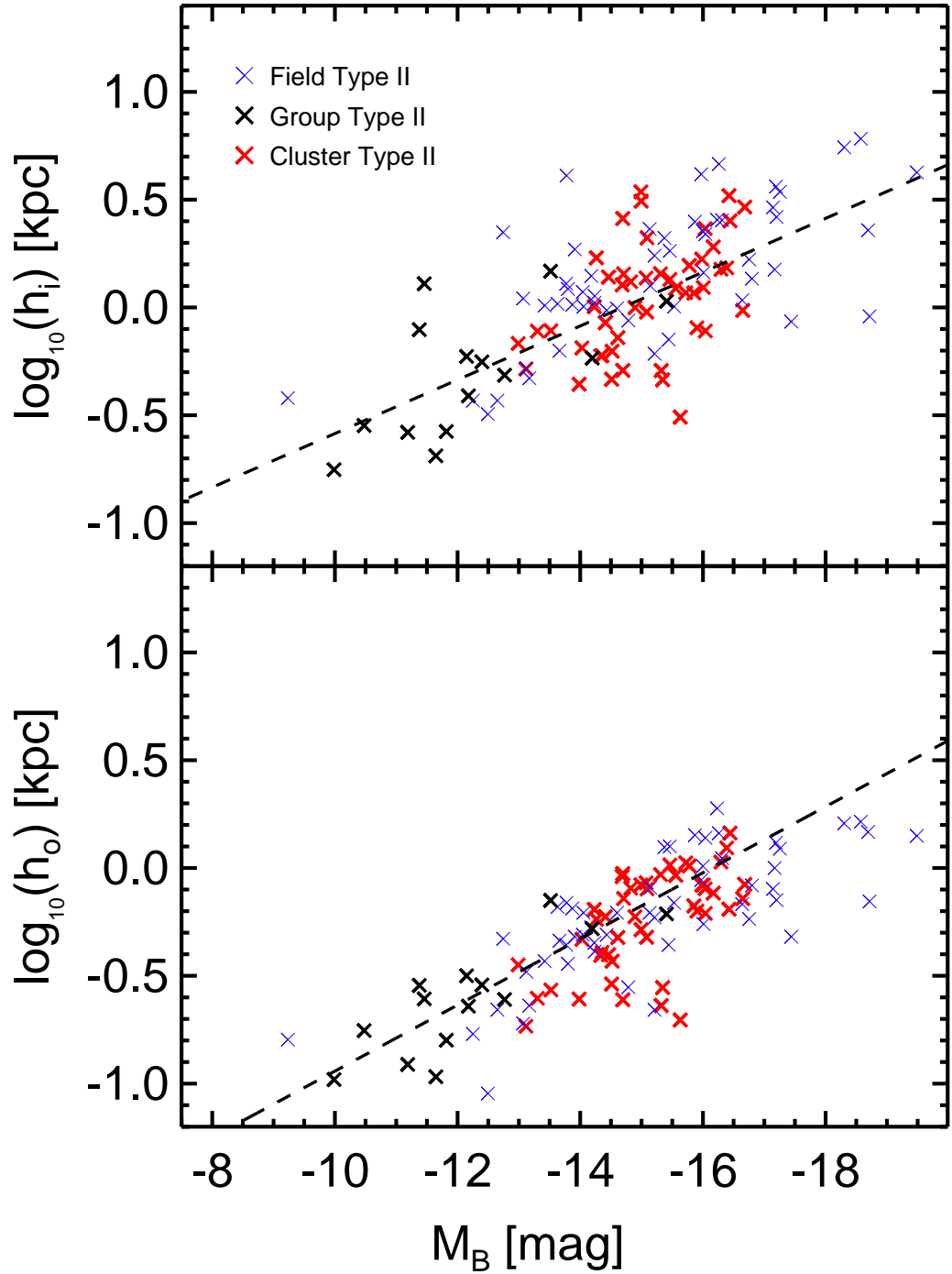}
\caption{Size distributions of Type II dwarfs as a function of environment. The upper and bottom panels show for inner and outer sizes, respectively. The dwarf galaxies in the field, group, and cluster are shown as blue, black, and red symbols, respectively. The dashed lines are the same dashed lines in Figure~\ref{fig2:sizemag}, which are measured from sizes of Type II dwarfs in the NGC 2784 group. \label{fig5:sizemagcomp_t2}}
\end{figure}

\begin{figure} 
\plotone{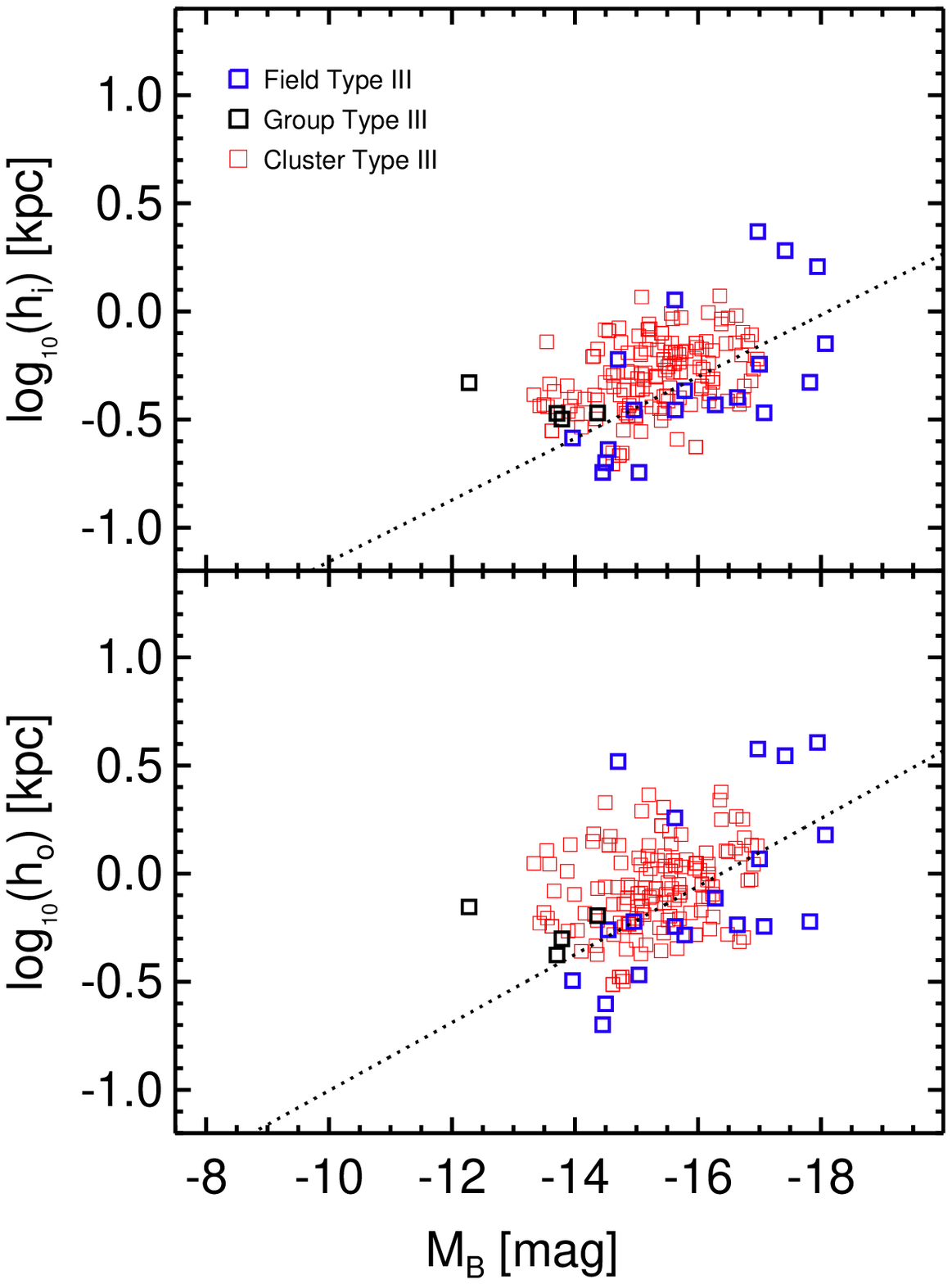}
\caption{Size distributions of Type III dwarfs as a function of environment. The upper and bottom panels show for inner and outer sizes, respectively. The dwarf galaxies in the field, group, and cluster are shown as blue, black, and red symbols, respectively. The dotted lines are measured together with sizes of field and group Type III dwarfs. \label{fig6:sizemagcomp_t3}}
\end{figure}

\ylee{In Figure~\ref{fig4:sizemagcomp_t1},~\ref{fig5:sizemagcomp_t2}, and ~\ref{fig6:sizemagcomp_t3}, we compare the sizes for the dwarfs in the field, group, and cluster. In this analysis, we do not distinguish early- and late-type dwarfs because they show similar size-magnitude relations (see Section~\ref{ssec:frequency}). When comparing size distributions, we use sizes normalized by the given size-magnitude relations of the dwarfs in the NGC 2784 group, which the relations for Type I, Type II, and Type III are shown as a solid line in Figure~\ref{fig4:sizemagcomp_t1}, a dashed line in Figure~\ref{fig5:sizemagcomp_t2}, and a dotted line in Figure~\ref{fig6:sizemagcomp_t3}, respectively. In Figure~\ref{fig4:sizemagcomp_t1}, the size distribution of group Type I dwarfs is almost identical to that for field and cluster Type I dwarfs ($P_{KS}=0.136$ for the field and $P_{KS}=0.480$ for the cluster). In Figure~\ref{fig5:sizemagcomp_t2}, inner and outer size distributions of group Type II dwarfs are also similar to those for field Type II dwarfs ($P_{KS}=0.118$ for inner size and $P_{KS}=0.109$ for outer size) and to those for cluster Type II dwarfs ($P_{KS}=0.711$ for inner size and $P_{KS}=0.346$ for outer size).} 

\ylee{Since we did not measure size-magitude relations of Type III dwarfs in the NGC 2784 group, we do not statistically compare the size distributions between the group and field. However, the size distributions of group Type III dwarfs seem to be located on that of field Type III dwarfs in Figure~\ref{fig6:sizemagcomp_t3}. For size-magnitude relations of Type III dwarfs, we measure these relations together with group and field dwarfs. The relations are followings :}
\begin{equation}
log_{10}(h_i) = -0.1468\times M_B -2.4151 \qquad\textrm{for inner size of Type III}
\end{equation}
\begin{equation}
log_{10}(h_o) = -0.1561\times M_B -2.5647 \qquad\textrm{for outer size of Type III}
\end{equation}
\ylee{The sizes of Type III dwarfs are compared using sizes normalized by Equation (5) and (6). The size distributions of cluster Type III dwarfs deviate from those of group and field Type III dwarfs ($P_{KS}=0.008$ for inner size and $P_{KS}=0.030$ for outer size). Sizes (median normalized sizes on a log scale, MNS, are 0.120 for inner size and 0.101 for outer size) of cluster Type III dwarfs are systematically larger than those of group and field Type III dwarfs (MNSs are -0.041 for inner size and -0.024 for outer size).}

\begin{table} 
\renewcommand{\thetable}{\arabic{table}}
\centering
\caption{Fractions of {\sbp} types with $-16$ mag $< M_B < -12$ mag} \label{tab:fraction2}
\begin{tabular}{c c c c}
\tablewidth{0pt}
\hline
\hline
\multicolumn{4}{c}{NGC 2784 group} \\
\hline
Morphology & Type I & Type II & Type III \\
\hline
\decimals
All (20)  &  9 (45.0$_{-10.6}^{+11.1}$\%)    & 7 (35.0$_{-9.7}^{+11.1}$\%) & 4 (20.0$_{-7.4}^{+10.3}$\%)     \\
Early-type (12) &  4 (33.3$_{-11.9}^{+14.4}$\%)   &  4 (33.3$_{-11.9}^{+14.4}$\%)  & 4 (33.3$_{-11.9}^{+14.4}$\%)     \\
Late-type (8) &  5 (62.5$_{-17.6}^{+14.8}$\%)   &   3 (37.5$_{-14.8}^{+17.6}$\%)   & 0 (0.0$_{-0.0}^{+11.0}$\%)  \\
Unknown (1) &  1 (100.0$_{-50.0}^{+0.0}$\%)   &   0 (0$_{-0.0}^{+50.0}$\%) & 0 (0$_{-0.0}^{+50.0}$\%) \\
\hline

\multicolumn{4}{c}{} \\
\hline
\hline
\multicolumn{4}{c}{Field} \\
\hline
Late-type (36) & 4 (11.1$_{-4.2}^{+6.3}$\%) &  23 (63.9$_{-8.3}^{+7.5}$\%) &  9 (25.0$_{-6.5}^{+7.8}$\%) \\
\hline

\multicolumn{4}{c}{} \\
\hline
\hline
\multicolumn{4}{c}{Virgo cluster} \\
\hline
All (327) & 187 (57.2$_{-2.8}^{+2.7}$\%) & 37 (11.3$_{-1.6}^{+1.9}$\%) & 103 (31.5$_{-2.5}^{+2.6}$\%) \\
Early-type (260) & 154 (59.2$_{-3.1}^{+3.0}$\%) & 18 (6.9$_{-1.4}^{+1.7}$\%) & 88 (33.8$_{-2.9}^{+3.0}$\%) \\
Late-type (67) & 33 (49.3$_{-6.1}^{+6.1}$\%) & 19 (28.4$_{-5.2}^{+5.8}$\%) & 15 (22.4$_{-4.7}^{+5.5}$\%) \\
\hline

\end{tabular}
\end{table}

\begin{table} 
\renewcommand{\thetable}{\arabic{table}}
\centering
\caption{Probabilities ($P_F$) of Fisher's Exact Test} \label{tab:Pfs}
\begin{tabular}{c c c}
\tablewidth{0pt}
\hline
\hline
Morphology (Environment) & All samples & $-16$ mag $<M_B<-12$ mag  \\
\hline
\decimals

All (Group vs. Field)   &  $5.23\times10^{-6}$    & $0.017$    \\
All (Group vs. Cluster)   &  $6.71\times10^{-5}$    & $0.021$    \\
All (Field vs. Cluster)   &  $2.20\times10^{-16}$    & $1.10\times10^{-12}$    \\
Early vs. Late (Group)   &  $0.107$    & $0.219$      \\
Early vs. Late (Cluster)  &  $2.31\times10^{-7}$   &  $2.52\times10^{-5}$       \\
Early (Group vs. Cluster)   &  $8.55\times10^{-5}$    &  $0.012$   \\
Late (Group vs. Cluster)   & $0.095$     & $0.379$    \\
Late (Group vs. Field)   &  $5.60\times10^{-6}$    & $0.009$    \\
Late (Field vs. Cluster)   &  $8.96\times10^{-8}$    & $1.37\times10^{-4}$    \\

\hline

\end{tabular}
\end{table}

\subsection{Comparison between sizes of the different {\sbp} types} \label{ssec:comptypes}

\begin{figure} 
\plotone{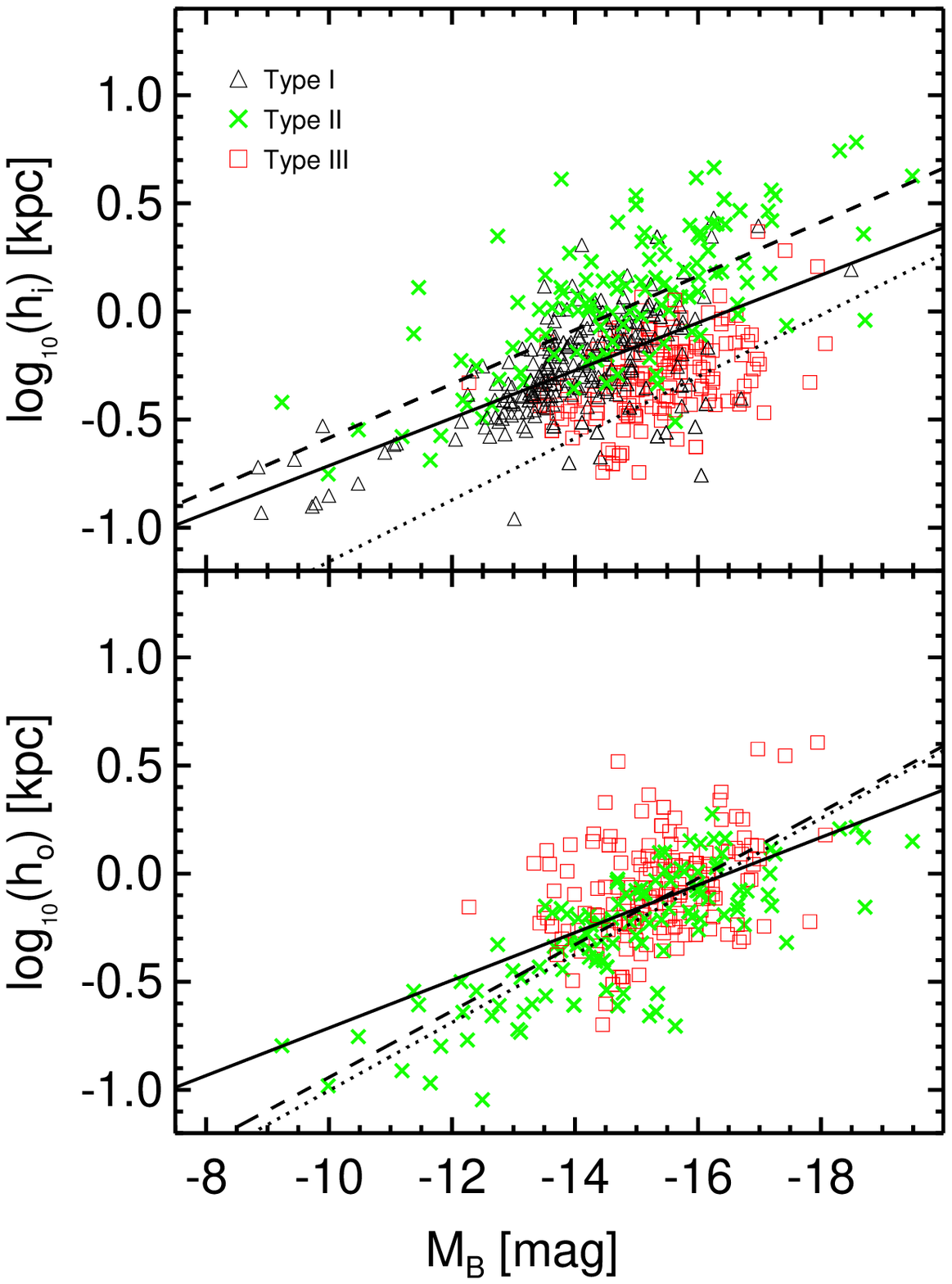}
\caption{Size distributions of dwarfs as a function of {\sbp} type. The upper and bottom panels show for inner and outer sizes, respectively. The Type I, Type II, and Type III dwarfs are shown as black trangles, green crosses, and red squares, respectively. The solid and dashed lines are size-magnitude relations for Type I and Type II dwarfs, respectively, which are the same lines in Figure~\ref{fig2:sizemag}. The dotted lines are size-magnitude relations for Type III dwarfs and are the same lines in Figure~\ref{fig6:sizemagcomp_t3}. \label{fig7:sizemagcomp_all}}
\end{figure}

\ylee{In the previous section, we compared size distributions of the {\sbp} type among environments. Size distributions of Type I and Type II dwarfs did not show environmental variations. Therefore, we do not divide the environments for comparing size distributions of {\sbp} types. Based on the sizes normalized by the size-magnitude relation of group Type I dwarfs (sold line in Figure~\ref{fig4:sizemagcomp_t1} and Figure~\ref{fig7:sizemagcomp_all}), the size distribution of Type I dwarfs is \yleem{smaller} than that of Type II dwarfs ($P_{KS}=2.53\times10^{-18}$ for inner sizes and $P_{KS} = 8.67\times10^{-5}$ for outer sizes) in Figure~\ref{fig7:sizemagcomp_all}. MNSs are 0.006 for a size of Type I dwarfs, 0.240 for an inner size of Type II dwarfs, and -0.084 for an outer size of Type II dwarfs.}

\ylee{Since cluster Type III dwarfs have somewhat larger sizes than group and field Type III dwarfs (see the previous section), we divide Type III dwarfs into Type III dwarfs in the field, group, and cluster (FGC) and in the field and group (FG) to consider the cases where cluster samples are included or not. Based on the sizes normalized by the size-magnitude relation of group Type I dwarfs, the inner size distributions of FGC and FG Type III dwarfs are smaller than that of Type I dwarfs in all environments ($P_{KS}= 1.71\times10^{-19}$ for FGC Type III and Type I; $P_{KS}=1.52\times10^{-7}$ for FG Type III and Type I). MNSs are -0.166 for FGC Type III and -0.302 for FG Type III (cf. MNS for Tyep I dwarfs in all environments is 0.006). The outer size distributions of FGC and GF Type III dwarfs are similar to that of Type I dwarfs in all environments ($P_{KS} = 0.058$ for FGC Type III and Type I; $P_{KS}=0.178$ for FG Type III and Type I).}

\ylee{Size distributions of Type II and Type III dwarfs are compared. Sizes normalized by the size-magnitude relations of group Type II dwarfs (dashed line in Figure~\ref{fig7:sizemagcomp_all}) are used. The inner sizes of FGC and FG Type III dwarfs are smaller than those of Type II dwarfs in all environments ($P_{KS} = 7.59\times10^{-34}$ for FGC Type III and Type II; $P_{KS}=6.47\times10^{-9}$ for FG Type III and Type II). MNSs are 0.037 for an inner size of Type II dwarfs, -0.374 for an inner size of FGC Type III dwarfs and -0.503 for an inner size of FG Type III dwarfs. The outer sizes of FGC Type III dwarfs are larger than those of Type II dwarfs in all environments ($P_{KS} = 1.18\times10^{-4}$ for FGC Type III and Type II). MNS is 0.048 for an outer size of FGC Type III dwarfs and MNS of an outer size of Type II dwarfs in all environments is -0.080. An outer size distribution of FG Type III dwarfs is similar to that of Type II in all environments ($P_{KS}=0.114$).}

\ylee{As a result, the inner sizes of Type II, Type I, and Type III dwarfs are in descending order of size. Outer size distributions of Type I and Type III dwarfs are similar and they are larger than those of Type II for FGC samples, although excluding cluster samples makes outer size distributions of Type III dwarfs similar to those of Type II dwarfs.}

\section{Discussion} \label{sec:summary}

\subsection{\ylee{Origins of {\sbp} types}}
\ylee{We study  the three {\sbp} types of dwarfs. For the origin of various {\sbp} types,} several processes such as the angular momentum distribution of the gas, a radial dropoff in efficient star formation, minor mergers, smooth gas accretion, external tidal interactions and scattering or migration of stars are suggested \citep{Mes63,Fre70,Sch04,Erw05,Deb06,Elm06,You07,Min12,Mun13,Lai14,Str17}. Out of these suggestions, scattering or migration of stars from the inner to outer region in a galaxy is the most popular explanation. For this process, bars/arms and clumps/holes are the main drivers to move the stars \citep{Lai14,Str17}. When considering that bars/arms are rare in less massive dwarfs, although they are frequently found in massive dwarfs \citep{Lis06,Jan14}, different {\sbp} types could be explained as scattering by clumps/holes rather than by bars/arms.

According to this study, there are variations of {\sbp} types among environments. \ylee{Common {\sbp} types of dwarfs in the field, group, and cluster are Type II, Type I and II, and Type I and III profiles, respectively.} These results seem to imply that the {\sbp} types can not be explained by only the internal processes and environmental dependent processes must be involved \citep{Erw12,Mal12,Roe12}. As a similar argument, \citet{Erw12} found different {\sbp} type fractions of S0 disk profiles in the field and the Virgo cluster and suggested that the {\sbp} types could be transformed by environmental effects in a cluster.

\subsection{\ylee{{\sbp} types of dwarfs in the field}}
Since the field environment has no or little influence on dwarfs, \ylee{late-type dwarfs are dominent galaxies \citep{Geh12}. They have three kinds of {\sbp} types. The type fractions do not change significantly as a function of luminosity (Figure~\ref{fig3:magdepen}). Out of the types, the most common {\sbp} types ($\sim$68\%) of them are Type II profiles (see Table~\ref{tab:fraction} and Table~\ref{tab:fraction2}). These results can be interpreted as that the internal processes to shape the {\sbp}s prefer to form Type II profiles rather than the other types (Type I and III) in the field environment.}

\subsection{\ylee{{\sbp} types of dwarfs in groups}}
According to the hierarchical structure formation scenario, field galaxies are assembled into groups of galaxies \citep{Ber09,deL12}. In this scenario, it is expected that the distributions of {\sbp} types in groups are similar to those in the field. \ylee{However, the distributions are different (see Table~\ref{tab:Pfs}). The distribution of {\sbp} types in group dwarfs with late-type morphologies is even different from that of field dwarfs. This seems to imply that environmental effects of groups can alter {\sbp} types before stellar populations or morphologies of field dwarfs infalling into the groups change. When group late-type dwarfs turn into group early-type dwarfs, the {\sbp} types are not significantly altered (see Table~\ref{tab:Pfs}).}

\ylee{In groups of galaxies, (wet) mergers and tidal interactions are the important environmental effects \citep{Moo96,Ver98,Mil04}.} These effects make the interstellar gases near the center of a perturbed galaxy move into the center of the galaxy to form stars \citep{Lak98,Moo96,Moo98}. In the outer region of the perturbed galaxy, the interactions induce gas losses and quenching of the star formation. \ylee{Furthermore, they make the outer disk re-distributed to have a larger size \citep{Pen09}. As a result of environmental effects, inner sizes of perturbed galaxies will become smaller while the outer sizes will be larger. Namely, some of group Type I dwarfs being smaller inner sizes and larger outer sizes than Type II dwarfs may be transformed from field Type II dwarfs. Therefore, the most common {\sbp} type of group dwarfs can be Type I and II profiles (see Table~\ref{tab:fraction} and Table~\ref{tab:fraction2}).
}

\subsection{\ylee{{\sbp} types of dwarfs in clusters}}

\ylee{Clusters of galaxies are formed through assembly of group galaxies and field galaxies \citep{Ber09,deL12}. Interestingly, the {\sbp} type distributions of all dwarfs in clusters are quite different from that in groups and in the field (see Table~\ref{tab:Pfs}). Furthermore, Type I and Type II fractions in clusters depend on luminosities of dwarfs (Figure~\ref{fig3:magdepen}). These seem to imply that difference of environmental mechanisms and strengths between in groups and clusters makes the {\sbp} types various. If dwarfs in clusters have similar environmental mechanism and strengths to those of groups, the {\sbp} type distributions might be similar. For example, in the case of cluster late-type dwarfs which may be infalling into the cluster for the first time and experience weak cluster environments (such as tidal interactions) at the outer part of the cluster, the {\sbp} type distribution of cluster late-type dwarfs is similar to that of group late-type dwarfs (see Table~\ref{tab:Pfs}).}

\ylee{Cluster early-type dwarfs show very different {\sbp} type distributions comparing with those of group early-type dwarfs. Cluster early-type dwarfs have a small Type II fraction ($\sim$8\%) and large fractions of the other {\sbp} types ($\sim$52\% for Type I and $\sim$40\% for Type III), while group early-type dwarfs have a small Type III fraction ($\sim$19\%) and large fractions of Type I ($\sim$38\%) and Type II ($\sim$43\%). Similarly, disk type fractions of lenticular galaxies (S0s) are also found in the Virgo cluster \citep{Erw12}, in which the disk type fractions of them are 45.8\% for Type I, 0\% for Type II, and 54.2\% for Type III. Furthermore, \citet{You07} roughly expects $\sim$40\%-50\% of Type III profiles in their merger-driven scenario. If cluster early-type dwarfs are descendants of cluster late-type dwarfs which are experiencing strong cluster environmental effects \citep{Bos08a,Bos08b}, the {\sbp} type distributions between cluster early- and late-type dwarfs can be interpreted as an evolutionary sequence. That is, most of the Type II profiles of cluster late-type dwarfs can transform to cluster early-type dwarfs with Type I or Type III profiles experiencing cluster environmental effects and this evolution of {\sbp} types can make populations of Type II profiles depleted.}

In clusters of galaxies, hydrodynamic interactions (i.e., ram pressure stripping), as well as gravitational interactions are the important environmental effects \citep{Bos08a,Smi13,Smi15}. \ylee{Since clusters have more massive galaxies and larger peculiar velocities of cluster galaxies than groups, harassment of gravitational interactions in clusters is stronger than in groups \citep{Moo96,Moo98}. It may boost transformations from Type II to Type I or Type III as well as transformations from Type I to Type III (see figure 7 of \citet{Moo99}), since cold gas funnels to the center of the galaxy and star formations are triggered.} Ram pressure between cold interstellar gas in galaxies infalling into the cluster and hot intracluster medium removes the cold gas in the galaxies. Since ram pressure is an inverse function of the surface stellar density of a galaxy \citep{Gun72}, it is easy to lose their gas in the outer region, while the inner region of the galaxy holds their gas against the pressure. Therefore, star formations can be kept in the central region of the galaxy and no star formation is maintained in the outer region \citep{Kro08}. 

\ylee{The gas-loss timescale due to ram pressure stripping is a function of the mass of dwarfs \citep{Mor00}. Massive dwarfs hold a gas in their center for a longer timescale than less massive dwarfs. This mass-depenent ram pressure stripping may result in the luminosity dependence of {\sbp} types which is found in Figure~\ref{fig3:magdepen}. That is, when low mass late-type dwarfs with Type II profiles fall into a cluster, the dwarfs experiencing tidal interactions transform into low mass late-type dwarfs with Type I profiles at the outer part of the cluster. Then, before transforming into the low mass early-type dwarfs with Type III profile, they completely lose their gas due to ram pressure stripping. Therefore, it is possible to have a large fraction of Type I dwarfs at the low luminosity regime.}

In this paper, we suggest that the different {\sbp} type fractions of dwarfs in the field, group, and cluster can be driven by combinations of environmental effects. However, the results we obtained depended heavily on the small number of groups and clusters. More data on groups and clusters in the future will help advance our understanding of the {\sbp}s of dwarfs and their transformations.

\acknowledgments

This research has made use of the KMTNet system operated by the Korea Astronomy and Space Science Institute (KASI) and the data were obtained at three host sites of CTIO in Chile, SAAO in South Africa, and SSO in Australia. DSM was supported in part by a Leading Edge Fund from the Canadian Foundation for Innovation (Project Number 30951) and a Discovery Grant from the Natural Sciences and Engineering Research Council of Canada.

\bibliography{kspgal}

\begin{thebibliography}{}
\expandafter\ifx\csname natexlab\endcsname\relax\def\natexlab#1{#1}\fi
\providecommand{\url}[1]{\href{#1}{#1}}

\bibitem[{{Aguerri} {et~al.}(2005){Aguerri}, {Iglesias-P{\'a}ramo},
  {V{\'{\i}}lchez}, {Mu{\~n}oz-Tu{\~n}{\'o}n}, \&
  {S{\'a}nchez-Janssen}}]{Agu05}
{Aguerri}, J.~A.~L., {Iglesias-P{\'a}ramo}, J., {V{\'{\i}}lchez}, J.~M.,
  {Mu{\~n}oz-Tu{\~n}{\'o}n}, C., \& {S{\'a}nchez-Janssen}, R. 2005, \aj, 130,
  475

\bibitem[{{Berrier} {et~al.}(2009){Berrier}, {Stewart}, {Bullock}, {Purcell},
  {Barton}, \& {Wechsler}}]{Ber09}
{Berrier}, J.~C., {Stewart}, K.~R., {Bullock}, J.~S., {et~al.} 2009, \apj, 690,
  1292

\bibitem[{{Binggeli} {et~al.}(1985){Binggeli}, {Sandage}, \& {Tammann}}]{Bin85}
{Binggeli}, B., {Sandage}, A., \& {Tammann}, G.~A. 1985, \aj, 90, 1681

\bibitem[{{Boller} {et~al.}(2016){Boller}, {Freyberg}, {Tr{\"u}mper}, {Haberl},
  {Voges}, \& {Nandra}}]{Bol16}
{Boller}, T., {Freyberg}, M.~J., {Tr{\"u}mper}, J., {et~al.} 2016, \aap, 588,
  A103

\bibitem[{{Boselli} {et~al.}(2008{\natexlab{a}}){Boselli}, {Boissier},
  {Cortese}, \& {Gavazzi}}]{Bos08a}
{Boselli}, A., {Boissier}, S., {Cortese}, L., \& {Gavazzi}, G.
  2008{\natexlab{a}}, \aap, 489, 1015

\bibitem[{{Boselli} {et~al.}(2008{\natexlab{b}}){Boselli}, {Boissier},
  {Cortese}, \& {Gavazzi}}]{Bos08b}
---. 2008{\natexlab{b}}, \apj, 674, 742

\bibitem[{{Boselli} {et~al.}(1997){Boselli}, {Tuffs}, {Gavazzi}, {Hippelein},
  \& {Pierini}}]{Bos97}
{Boselli}, A., {Tuffs}, R.~J., {Gavazzi}, G., {Hippelein}, H., \& {Pierini}, D.
  1997, \aaps, 121, 507

\bibitem[{{De Lucia} {et~al.}(2012){De Lucia}, {Weinmann}, {Poggianti},
  {Arag{\'o}n-Salamanca}, \& {Zaritsky}}]{deL12}
{De Lucia}, G., {Weinmann}, S., {Poggianti}, B.~M., {Arag{\'o}n-Salamanca}, A.,
  \& {Zaritsky}, D. 2012, \mnras, 423, 1277

\bibitem[{{Debattista} {et~al.}(2006){Debattista}, {Mayer}, {Carollo}, {Moore},
  {Wadsley}, \& {Quinn}}]{Deb06}
{Debattista}, V.~P., {Mayer}, L., {Carollo}, C.~M., {et~al.} 2006, \apj, 645,
  209

\bibitem[{{Elmegreen} \& {Hunter}(2006)}]{Elm06}
{Elmegreen}, B.~G., \& {Hunter}, D.~A. 2006, \apj, 636, 712

\bibitem[{{Erwin} {et~al.}(2005){Erwin}, {Beckman}, \& {Pohlen}}]{Erw05}
{Erwin}, P., {Beckman}, J.~E., \& {Pohlen}, M. 2005, \apjl, 626, L81

\bibitem[{{Erwin} {et~al.}(2012){Erwin}, {Guti{\'e}rrez}, \& {Beckman}}]{Erw12}
{Erwin}, P., {Guti{\'e}rrez}, L., \& {Beckman}, J.~E. 2012, \apjl, 744, L11

\bibitem[{{Erwin} {et~al.}(2008){Erwin}, {Pohlen}, \& {Beckman}}]{Erw08}
{Erwin}, P., {Pohlen}, M., \& {Beckman}, J.~E. 2008, \aj, 135, 20

\bibitem[{{Fern{\'a}ndez Lorenzo} {et~al.}(2012){Fern{\'a}ndez Lorenzo},
  {Sulentic}, {Verdes-Montenegro}, {Ruiz}, {Sabater}, \& {S{\'a}nchez}}]{Fer12}
{Fern{\'a}ndez Lorenzo}, M., {Sulentic}, J., {Verdes-Montenegro}, L., {et~al.}
  2012, \aap, 540, A47

\bibitem[{{Fouque} {et~al.}(1992){Fouque}, {Gourgoulhon}, {Chamaraux}, \&
  {Paturel}}]{Fou92}
{Fouque}, P., {Gourgoulhon}, E., {Chamaraux}, P., \& {Paturel}, G. 1992, \aaps,
  93, 211

\bibitem[{{Freeman}(1970)}]{Fre70}
{Freeman}, K.~C. 1970, \apj, 160, 811

\bibitem[{{Garcia}(1993)}]{Gar93}
{Garcia}, A.~M. 1993, \aaps, 100, 47

\bibitem[{{Gavazzi} {et~al.}(2010){Gavazzi}, {Fumagalli}, {Cucciati}, \&
  {Boselli}}]{Gav10}
{Gavazzi}, G., {Fumagalli}, M., {Cucciati}, O., \& {Boselli}, A. 2010, \aap,
  517, A73

\bibitem[{{Geha} {et~al.}(2012){Geha}, {Blanton}, {Yan}, \& {Tinker}}]{Geh12}
{Geha}, M., {Blanton}, M.~R., {Yan}, R., \& {Tinker}, J.~L. 2012, \apj, 757, 85

\bibitem[{{Gunn} \& {Gott}(1972)}]{Gun72}
{Gunn}, J.~E., \& {Gott}, III, J.~R. 1972, \apj, 176, 1

\bibitem[{{Herrmann} {et~al.}(2013){Herrmann}, {Hunter}, \&
  {Elmegreen}}]{Her13}
{Herrmann}, K.~A., {Hunter}, D.~A., \& {Elmegreen}, B.~G. 2013, \aj, 146, 104

\bibitem[{{Hunter} \& {Elmegreen}(2004)}]{Hun04}
{Hunter}, D.~A., \& {Elmegreen}, B.~G. 2004, \aj, 128, 2170

\bibitem[{{Hunter} \& {Elmegreen}(2006)}]{Hun06}
---. 2006, \apjs, 162, 49

\bibitem[{{Jansen} {et~al.}(2000){Jansen}, {Franx}, {Fabricant}, \&
  {Caldwell}}]{Jan00}
{Jansen}, R.~A., {Franx}, M., {Fabricant}, D., \& {Caldwell}, N. 2000, \apjs,
  126, 271

\bibitem[{{Janz} {et~al.}(2016){Janz}, {Laurikainen}, {Laine}, {Salo}, \&
  {Lisker}}]{Jan16}
{Janz}, J., {Laurikainen}, E., {Laine}, J., {Salo}, H., \& {Lisker}, T. 2016,
  \mnras, 461, L82

\bibitem[{{Janz} {et~al.}(2012){Janz}, {Laurikainen}, {Lisker}, {Salo},
  {Peletier}, {Niemi}, {den Brok}, {Toloba}, {Falc{\'o}n-Barroso}, {Boselli},
  \& {Hensler}}]{Jan12}
{Janz}, J., {Laurikainen}, E., {Lisker}, T., {et~al.} 2012, \apjl, 745, L24

\bibitem[{{Janz} {et~al.}(2014){Janz}, {Laurikainen}, {Lisker}, {Salo},
  {Peletier}, {Niemi}, {Toloba}, {Hensler}, {Falc{\'o}n-Barroso}, {Boselli},
  {den Brok}, {Hansson}, {Meyer}, {Ry{\'s}}, \& {Paudel}}]{Jan14}
---. 2014, \apj, 786, 105

\bibitem[{{Kim} {et~al.}(2014){Kim}, {Rey}, {Jerjen}, {Lisker}, {Sung}, {Lee},
  {Chung}, {Pak}, {Yi}, \& {Lee}}]{Kim14}
{Kim}, S., {Rey}, S.-C., {Jerjen}, H., {et~al.} 2014, \apjs, 215, 22

\bibitem[{{Kim} {et~al.}(2016){Kim}, {Lee}, {Park}, {Kim}, {Cha}, {Lee}, {Han},
  {Chun}, \& {Yuk}}]{Kim16}
{Kim}, S.-L., {Lee}, C.-U., {Park}, B.-G., {et~al.} 2016, Journal of the Korean
  Astronomical Society, 49, 37

\bibitem[{{Kormendy}(1985)}]{Kor85}
{Kormendy}, J. 1985, \apjl, 292, L9

\bibitem[{{Kormendy} \& {Bender}(2012)}]{Kor12}
{Kormendy}, J., \& {Bender}, R. 2012, \apjs, 198, 2

\bibitem[{{Kronberger} {et~al.}(2008){Kronberger}, {Kapferer}, {Ferrari},
  {Unterguggenberger}, \& {Schindler}}]{Kro08}
{Kronberger}, T., {Kapferer}, W., {Ferrari}, C., {Unterguggenberger}, S., \&
  {Schindler}, S. 2008, \aap, 481, 337

\bibitem[{{Laine} {et~al.}(2014){Laine}, {Laurikainen}, {Salo}, {Comer{\'o}n},
  {Buta}, {Zaritsky}, {Athanassoula}, {Bosma}, {Mu{\~n}oz-Mateos}, {Gadotti},
  {Hinz}, {Erroz-Ferrer}, {Gil de Paz}, {Kim}, {Men{\'e}ndez-Delmestre},
  {Mizusawa}, {Regan}, {Seibert}, \& {Sheth}}]{Lai14}
{Laine}, J., {Laurikainen}, E., {Salo}, H., {et~al.} 2014, \mnras, 441, 1992

\bibitem[{{Lake} {et~al.}(1998){Lake}, {Katz}, \& {Moore}}]{Lak98}
{Lake}, G., {Katz}, N., \& {Moore}, B. 1998, \apj, 495, 152

\bibitem[{{Lisker} {et~al.}(2006){Lisker}, {Grebel}, \& {Binggeli}}]{Lis06}
{Lisker}, T., {Grebel}, E.~K., \& {Binggeli}, B. 2006, \aj, 132, 497

\bibitem[{{Makarov} \& {Karachentsev}(2011)}]{Mak11}
{Makarov}, D., \& {Karachentsev}, I. 2011, \mnras, 412, 2498

\bibitem[{{Maltby} {et~al.}(2012){Maltby}, {Gray}, {Arag{\'o}n-Salamanca},
  {Wolf}, {Bell}, {Jogee}, {H{\"a}u{\ss}ler}, {Barazza}, {B{\"o}hm}, \&
  {Jahnke}}]{Mal12}
{Maltby}, D.~T., {Gray}, M.~E., {Arag{\'o}n-Salamanca}, A., {et~al.} 2012,
  \mnras, 419, 669

\bibitem[{{McDonald} {et~al.}(2011){McDonald}, {Courteau}, {Tully}, \&
  {Roediger}}]{McD11}
{McDonald}, M., {Courteau}, S., {Tully}, R.~B., \& {Roediger}, J. 2011, \mnras,
  414, 2055

\bibitem[{{Mei} {et~al.}(2007){Mei}, {Blakeslee}, {C{\^o}t{\'e}}, {Tonry},
  {West}, {Ferrarese}, {Jord{\'a}n}, {Peng}, {Anthony}, \& {Merritt}}]{Mei07}
{Mei}, S., {Blakeslee}, J.~P., {C{\^o}t{\'e}}, P., {et~al.} 2007, \apj, 655,
  144

\bibitem[{{Mestel}(1963)}]{Mes63}
{Mestel}, L. 1963, \mnras, 126, 553

\bibitem[{{Meyer} {et~al.}(2014){Meyer}, {Lisker}, {Janz}, \&
  {Papaderos}}]{Mey14}
{Meyer}, H.~T., {Lisker}, T., {Janz}, J., \& {Papaderos}, P. 2014, \aap, 562,
  A49

\bibitem[{{Miles} {et~al.}(2004){Miles}, {Raychaudhury}, {Forbes},
  {Goudfrooij}, {Ponman}, \& {Kozhurina-Platais}}]{Mil04}
{Miles}, T.~A., {Raychaudhury}, S., {Forbes}, D.~A., {et~al.} 2004, \mnras,
  355, 785

\bibitem[{{Minchev} {et~al.}(2012){Minchev}, {Famaey}, {Quillen}, {Di Matteo},
  {Combes}, {Vlaji{\'c}}, {Erwin}, \& {Bland-Hawthorn}}]{Min12}
{Minchev}, I., {Famaey}, B., {Quillen}, A.~C., {et~al.} 2012, \aap, 548, A126

\bibitem[{{Moon} {et~al.}(2016){Moon}, {Kim}, {Lee}, {Pak}, {Park}, {He},
  {Antoniadis}, {Ni}, {Lee}, {Kim}, {Park}, {Kim}, {Cha}, {Lee}, \&
  {Gonzalez}}]{Moo16}
{Moon}, D.-S., {Kim}, S.~C., {Lee}, J.-J., {et~al.} 2016, in \procspie, Vol.
  9906, Society of Photo-Optical Instrumentation Engineers (SPIE) Conference
  Series, 99064I

\bibitem[{{Moore} {et~al.}(1996){Moore}, {Katz}, {Lake}, {Dressler}, \&
  {Oemler}}]{Moo96}
{Moore}, B., {Katz}, N., {Lake}, G., {Dressler}, A., \& {Oemler}, A. 1996,
  \nat, 379, 613

\bibitem[{{Moore} {et~al.}(1998){Moore}, {Lake}, \& {Katz}}]{Moo98}
{Moore}, B., {Lake}, G., \& {Katz}, N. 1998, \apj, 495, 139

\bibitem[{{Moore} {et~al.}(1999){Moore}, {Lake}, {Quinn}, \& {Stadel}}]{Moo99}
{Moore}, B., {Lake}, G., {Quinn}, T., \& {Stadel}, J. 1999, \mnras, 304, 465

\bibitem[{{Mori} \& {Burkert}(2000)}]{Mor00}
{Mori}, M., \& {Burkert}, A. 2000, \apj, 538, 559

\bibitem[{{Mu{\~n}oz-Mateos} {et~al.}(2013){Mu{\~n}oz-Mateos}, {Sheth}, {Gil de
  Paz}, {Meidt}, {Athanassoula}, {Bosma}, {Comer{\'o}n}, {Elmegreen},
  {Elmegreen}, {Erroz-Ferrer}, {Gadotti}, {Hinz}, {Ho}, {Holwerda}, {Jarrett},
  {Kim}, {Knapen}, {Laine}, {Laurikainen}, {Madore}, {Menendez-Delmestre},
  {Mizusawa}, {Regan}, {Salo}, {Schinnerer}, {Seibert}, {Skibba}, \&
  {Zaritsky}}]{Mun13}
{Mu{\~n}oz-Mateos}, J.~C., {Sheth}, K., {Gil de Paz}, A., {et~al.} 2013, \apj,
  771, 59

\bibitem[{{Pak} {et~al.}(2014){Pak}, {Rey}, {Lisker}, {Lee}, {Kim}, {Sung},
  {Jerjen}, \& {Chung}}]{Pak14}
{Pak}, M., {Rey}, S.-C., {Lisker}, T., {et~al.} 2014, \mnras, 445, 630

\bibitem[{{Park} {et~al.}(2017){Park}, {Moon}, {Zaritsky}, {Pak}, {Lee}, {Kim},
  {Kim}, \& {Cha}}]{Par17}
{Park}, H.~S., {Moon}, D.-S., {Zaritsky}, D., {et~al.} 2017, \apj, 848, 19

\bibitem[{{Paturel} {et~al.}(2003){Paturel}, {Petit}, {Prugniel}, {Theureau},
  {Rousseau}, {Brouty}, {Dubois}, \& {Cambr{\'e}sy}}]{Pat03}
{Paturel}, G., {Petit}, C., {Prugniel}, P., {et~al.} 2003, \aap, 412, 45

\bibitem[{{Pe{\~n}arrubia} {et~al.}(2009){Pe{\~n}arrubia}, {Navarro},
  {McConnachie}, \& {Martin}}]{Pen09}
{Pe{\~n}arrubia}, J., {Navarro}, J.~F., {McConnachie}, A.~W., \& {Martin},
  N.~F. 2009, \apj, 698, 222

\bibitem[{{Roediger} {et~al.}(2012){Roediger}, {Courteau},
  {S{\'a}nchez-Bl{\'a}zquez}, \& {McDonald}}]{Roe12}
{Roediger}, J.~C., {Courteau}, S., {S{\'a}nchez-Bl{\'a}zquez}, P., \&
  {McDonald}, M. 2012, \apj, 758, 41

\bibitem[{{Sandage} \& {Binggeli}(1984)}]{San84}
{Sandage}, A., \& {Binggeli}, B. 1984, \aj, 89, 919

\bibitem[{{Schaye}(2004)}]{Sch04}
{Schaye}, J. 2004, \apj, 609, 667

\bibitem[{{Smith} {et~al.}(2013){Smith}, {Duc}, {Candlish}, {Fellhauer},
  {Sheen}, \& {Gibson}}]{Smi13}
{Smith}, R., {Duc}, P.~A., {Candlish}, G.~N., {et~al.} 2013, \mnras, 436, 839

\bibitem[{{Smith} {et~al.}(2015){Smith}, {S{\'a}nchez-Janssen}, {Beasley},
  {Candlish}, {Gibson}, {Puzia}, {Janz}, {Knebe}, {Aguerri}, {Lisker},
  {Hensler}, {Fellhauer}, {Ferrarese}, \& {Yi}}]{Smi15}
{Smith}, R., {S{\'a}nchez-Janssen}, R., {Beasley}, M.~A., {et~al.} 2015,
  \mnras, 454, 2502

\bibitem[{{Struck} \& {Elmegreen}(2017)}]{Str17}
{Struck}, C., \& {Elmegreen}, B.~G. 2017, \mnras, 469, 1157

\bibitem[{{Tonry} {et~al.}(2001){Tonry}, {Dressler}, {Blakeslee}, {Ajhar},
  {Fletcher}, {Luppino}, {Metzger}, \& {Moore}}]{Ton01}
{Tonry}, J.~L., {Dressler}, A., {Blakeslee}, J.~P., {et~al.} 2001, \apj, 546,
  681

\bibitem[{{Tully}(1987)}]{Tul87}
{Tully}, R.~B. 1987, \apj, 321, 280

\bibitem[{{Verdes-Montenegro} {et~al.}(1998){Verdes-Montenegro}, {Yun},
  {Perea}, {del Olmo}, \& {Ho}}]{Ver98}
{Verdes-Montenegro}, L., {Yun}, M.~S., {Perea}, J., {del Olmo}, A., \& {Ho},
  P.~T.~P. 1998, \apj, 497, 89

\bibitem[{{Wilson}(1927)}]{Wil27}
{Wilson}, E.~B. 1927, J. Amerc. Statist. Assoc., 22, 209

\bibitem[{{Young} {et~al.}(2014){Young}, {Jerjen}, {L{\'o}pez-S{\'a}nchez}, \&
  {Koribalski}}]{You14}
{Young}, T., {Jerjen}, H., {L{\'o}pez-S{\'a}nchez}, {\'A}.~R., \& {Koribalski},
  B.~S. 2014, \mnras, 444, 3052

\bibitem[{{Younger} {et~al.}(2007){Younger}, {Cox}, {Seth}, \&
  {Hernquist}}]{You07}
{Younger}, J.~D., {Cox}, T.~J., {Seth}, A.~C., \& {Hernquist}, L. 2007, \apj,
  670, 269

\end{thebibliography}

\end{document}